\documentclass[12pt]{article}
\usepackage{array}
\usepackage{graphicx}
\usepackage{amssymb}
\usepackage{amsmath}
\input{epsf}
\usepackage{cite}
\def\@fmsl@sh#1#2#3{\m@th\ooalign{$\hfil#1\mkern#2/\hfil$\crcr$#1#3$}}
 \def\eq#1\en{\begin{equation}#1\end{equation}}
\def\s[#1,#2]{[#1\stackrel{\star}{,}#2]}
\def\sx[#1,#2]{[#1\stackrel{\star_{x}}{,}#2]}

\newcommand{\nc}{\newcommand}
\nc{\beq}{\begin{equation}}
\nc{\eeq}{\end{equation}}
\nc{\beqa}{\begin{eqnarray}}
\nc{\eeqa}{\end{eqnarray}}

\def\bc{\begin{center}}
\def\ec{\end{center}}

\def\to{\rightarrow}

\def\gsim{\mathrel{\mathpalette\atversim>}}

\def\bc{\begin{center}}
\def\ec{\end{center}}

\def\gsim{\mathrel{\rlap{\lower4pt\hbox{\hskip1pt$\sim$}}

    \raise1pt\hbox{$>$}}}       

\def\gsim{\mathrel{\rlap{\lower4pt\hbox{\hskip1pt$\sim$}}
    \raise1pt\hbox{$>$}}}       

\begin{document}
\makeatletter
\def\fmslash{\@ifnextchar[{\fmsl@sh}{\fmsl@sh[0mu]}}
\def\fmsl@sh[#1]#2{%
  \mathchoice
    {\@fmsl@sh\displaystyle{#1}{#2}}%
    {\@fmsl@sh\textstyle{#1}{#2}}%
    {\@fmsl@sh\scriptstyle{#1}{#2}}%
    {\@fmsl@sh\scriptscriptstyle{#1}{#2}}}
\def\@fmsl@sh#1#2#3{\m@th\ooalign{$\hfil#1\mkern#2/\hfil$\crcr$#1#3$}}
\makeatother

\thispagestyle{empty}
\begin{titlepage}
\boldmath
\begin{center}
  \Large {\bf On the unitarity of linearized General Relativity coupled to matter}
  \end{center}
\unboldmath
\vspace{0.2cm}
\begin{center}
{{\large Michael Atkins}\footnote{m.atkins@sussex.ac.uk} and 
{\large Xavier Calmet}\footnote{x.calmet@sussex.ac.uk}
}
 \end{center}
\begin{center}
{\sl Physics and Astronomy, 
University of Sussex,  \\ Falmer, Brighton, BN1 9QH, UK 
}
\end{center}
\vspace{\fill}
\begin{abstract}
\noindent
We consider the unitarity of the S-matrix for linearized General Relativity coupled to particle physics models. Taking renormalization group effects of the Planck mass into account, we find that the scale at which unitarity  is violated is strongly dependent on the particle content of the theory. We find that the requirement that the S-matrix be unitary up to the scale at which quantum gravitational effects become strong implies a bound on the particle content of the model.

\end{abstract}  
\end{titlepage}



\newpage
General Relativity is remarkably successful on macroscopic scales and it describes all observations and experiments performed on distances from cosmological scales to distances of 10 $\mu$m, see e.g. \cite{review} for a review. More experiments are planned to probe General Relativity on yet shorter scales studying deviations of Newton's potential while the Large Hadron Collider will probe gravity in the TeV region \cite{Calmet:2009gn}. Within conventional physics, one does not expect deviations of General Relativity before one reaches energy scales close to the Planck scale or some $10^{19} \ \mbox{GeV}$. One expects that at this energy scale, quantum gravitational effects will become relevant. However, it is notoriously difficult to make sense of General Relativity once  second quantization effects are taken into account.  In particular the quantum field theory obtained by linearizing the Einstein-Hilbert action is not renormalizable, at least in a perturbative manner.  Nevertheless, General Relativity at the quantum level can be treated as an effective field theory (see e.g. \cite{Donoghue:1993eb}). In this letter we shall focus on the the coupling of gravity to matter and we will investigate whether the corresponding linearized theory leads to a unitary S-matrix. 

A similar study has already been performed by Han and Willenbrock \cite{Han:2004wt}. Although we agree with their calculations for the tree level amplitudes, we shall push the discussion further taking the renormalization group evolution of Newton's constant into account which turns out to be crucial in order to interpret the results correctly. We point out that it is important to compare the scale at which unitarity is violated to the scale at which quantum gravity effects become strong. The true scale for strong quantum gravitational effects can be determined dynamically using the renormalization group equation of Newton's constant. This enables us to derive a bound on the particle content of the particle physics model coupled to linearized General Relativity. The consequences for  these models are discussed.

We shall start from the usual four dimensional  Einstein-Hilbert action coupled to real scalar fields, Weyl fermions and vector fields treating them as massless particles
\begin{eqnarray} \label{effac1}
S[g,\phi, \psi, A_\mu]&= -\int d^4x \sqrt{-\det(g)} & \left   (\frac{1}{16\pi G_N} R+  \frac{1}{2} g^{\mu\nu} \partial_\mu \phi 
 \partial_\nu \phi + \xi R \phi^2 + \right .  \\ && \nonumber \left . +  e \bar \psi i \gamma^\mu D_\mu \psi + \frac{1}{4} F_{\mu\nu}F^{\mu\nu} \right)
\end{eqnarray}
where $e$ is the tetrad, $D_\mu=\partial_\mu + w^{ab}_\mu \sigma_{ab}/2$ and $ w^{ab}_\mu$ is the spin connection which can be expressed in terms of the tetrad.
This action can be linearized using  $g_{\mu\nu}=\eta_{\mu\nu}+\sqrt{2}h_{\mu\nu}/\bar M_P + {\cal O}(\bar M_P^{-2})$, where the scale, i.e the reduced Planck mass, appearing in this expansion is fixed by the requirement that the kinetic term of the graviton be  canonically normalized. One obtains the following Lagrangian
\begin{eqnarray} \label{efflag1}
L= -\frac{1}{4} h^{\mu\nu} \square h_{\mu\nu} +\frac{1}{4} h \square h -\frac{1}{2} h^{\mu\nu} \partial_\mu \partial_\nu h +   \frac{1}{2}h^{\mu\nu} \partial_\mu \partial_\alpha h_\nu^\alpha - \frac{\sqrt{2}}{\bar M_P} h^{\mu\nu}  T_{\mu\nu} + {\cal O}( \bar M^{-2}_P)
\end{eqnarray}
where $T^{\mu\nu}$ is the energy-momentum tensor corresponding to the matter content of the theory. This action can be regarded as an effective action valid up to $\bar M_P\sim 2.43 \times 10^{18} \ \mbox{GeV}$.  Traditionally one expects that gravitational interactions become strong above this energy scale and  the metric should not be linearizable at higher energies. In that sense we can consider that linearized General Relativity is an effective theory valid up to an energy scale corresponding to the reduced Planck mass.

One of the key consistency checks for an effective theory is that its S-matrix be unitary up to the scale where its description is assumed to  hold. As in  \cite{Han:2004wt}, we study the gravitational scattering at tree level of the real scalars $s$, Weyl fermions $\psi$ and vector bosons $V$ included in the particle physics model under consideration.  As they have done we restrict ourselves to the case where initial and final states consist of different particles. This simplifies the calculations tremendously since only s-channels need to be considered. We have calculated the amplitudes for the different processes. Our results can be found in table \ref{t1} and agree with those obtained  in \cite{Han:2004wt}. 
\begin{table*}[tbh]
\resizebox{\textwidth}{!}{
\begin{tabular}{|c|c|c|c|c|c|}  
\hline
 $\to$ & $s' \bar s'$ & $\psi'_+\bar \psi'_- $ & $ \psi'_-\bar \psi'_+ $ & $V'_+ V'_-$ & $V'_- V'_+$ \\
\hline $s \bar s$  & $-2\pi G_N s({1}/{3}d^2_{0,0} -
{1}/{3}(1+12\xi)^2d^0_{0,0})$ & $ -2\pi G_N s \sqrt{1/3}\ d^2_{0,1}$
& $ -2\pi G_N s \sqrt{1/3}\ d^2_{0,-1}$  & $-4\pi G_N s \sqrt{1/3}\ d^2_{0,2} $  & $ -4\pi G_N s\sqrt{1/3}\ d^2_{0,-2} $\\
\hline $\psi_+\bar \psi_- $ &
                $ -2\pi G_N s \sqrt{1/3}\ d^2_{1,0}$ & $ -2\pi G_N s d^2_{1,1} $
& $  -2\pi G_N sd^2_{1,-1} $ &  $-4\pi G_N s\ d^2_{1,2}$ & $-4\pi G_N s \  d^2_{1,-2}$ \\
\hline $\psi_-\bar \psi_+ $ &
                $-2\pi G_N s \sqrt{1/3}\ d^2_{-1,0}$ & $ -2\pi G_N s d^2_{-1,1} $
& $ -2\pi G_N s d^2_{-1,-1} $ &   $ -4\pi G_N s 2\ d^2_{-1,2}$ & $ -4\pi G_N s 2\ d^2_{-1,-2}$ \\
\hline $V_+ V_- $ &  $ -4\pi G_N s \sqrt{1/3}\ d^2_{2,0}$ &  $ -4\pi G_N s\
d^2_{2,1}$ &$
 -4\pi G_N s\ d^2_{2,-1}  $ & $-8\pi G_N s\ d^2_{2,2}$ & $-8\pi G_N s\ d^2_{2,-2}$ \\
\hline $V_- V_+  $ &  $-4\pi G_N s\sqrt{1/3}\ d^2_{-2,0}$ &  $ -4\pi G_N s\
d^2_{-2,1}$ &$
 -4\pi G_N s\ d^2_{-2,-1}  $ & $-8\pi G_N s\ d^2_{-2,2}$ & $-8\pi G_N s\ d^2_{-2,-2}$ \\
\hline  
\end{tabular}}
\caption{Scattering amplitudes for real scalars, fermions, and vector
bosons via s-channel graviton exchange in terms of the Wigner
$d$ functions \cite{Amsler:2008zzb}.  $G_N=1/M_P^2$ is Newton's constant and $s=E^2_{\rm{CM}}$ is the center of mass energy squared. We have used the helicity basis as in  \cite{Han:2004wt}}  \label{t1}
\end{table*}
The partial wave amplitudes $a_J$ can be determined using ${\cal A} =16 \pi \sum_J (2J+1) a_J d^J_{\mu,\mu^\prime}$.  As  is well known, the S-matrix is unitary if  $| \mbox{Re} \ a_J|\le1/2$. Looking at the $J=0$ amplitude we can deduce the first new result. If we request that the effective action remains valid up to the reduced Planck mass, i.e. we set  $\sqrt{s}=\bar M_P$, we obtain the following bound on the non-minimal coupling of the scalar field to the Ricci scalar:  
\begin{eqnarray} \label{boundxi}
-\frac{ 4\sqrt{6 \pi N_S}  +N_S }{12 N_S} \le \xi \le \frac{ 4\sqrt{6 \pi N_S}  -N_S }{12 N_S}.
\end{eqnarray}
In the standard model there is one Higgs doublet and hence four real scalars. Thus $N_S=4$ and we find  $-0.81\le \xi \le 0.64$ numerically. Note that the conformal value, $\xi=-1/12=-0.083$, is within this range and that in the limit $N_S \to \infty$, $\xi$ is forced to take the conformal value. If the model under consideration is to be valid up to the reduced Planck mass, the parameter $\xi$ needs to be rather small and is theoretically very tightly constrained. Clearly this casts some serious doubts on the validity of certain inflationary models such as, for example, the model proposed in \cite{Bezrukov:2007ep} where the Higgs boson plays the role of the inflaton, see also \cite{Barbon:2009ya} where a similar  observation was made.  The model  \cite{Bezrukov:2007ep}, although beautiful and minimalistic, requires some new physics below the reduced Planck mass to fix the unitarity problem.

Let us now look at the $J=2$ partial wave. Using the standard trick, as done in \cite{Han:2004wt,Lee:1977eg}, we can apply the unitarity bound to the scattering of a superposition of states. The  $J=2$ partial wave amplitude for the gravitational scattering of a state $|\sqrt{1/3} \sum s  s + \sum \psi_- \bar \psi_++ 2 \sum V V \rangle$ is given by
\begin{eqnarray} 
a_2=-\frac{1}{320 \pi} \frac{s}{\bar M_P^2} N
\end{eqnarray}
with $N=1/3 N_S + N_\psi+4 N_V$, where $N_S$, $N_\psi$ and $N_V$ are respectively the number of real scalar fields, Weyl fermions and vector bosons in the model under consideration.  The unitarity bound $| \mbox{Re} \ a_2|\le1/2$ implies a violation of unitarity of the S-matrix for center of mass energies $E_{\rm{CM}} > \bar M_P \sqrt{160 \pi/N}$.  

Naively, it seems that in particle physics models with a large number of fields the unitarity  of the S-matrix could be violated at energies below the reduced Planck mass. However, there is a physical effect which has not been included in  \cite{Han:2004wt} which has deep consequences for this study. It has been pointed out \cite{Calmet:2008tn,Calmet:2008df} that a large number of fields in a particle physics model can lead to a sizable running of the Planck mass. The $N$ fields introduced in the theory will renormalize the graviton propagator. The renormalization group equation for the reduced  Planck mass reads \cite{Calmet:2008tn,Larsen:1995ax,Kabat:1995eq,Vassilevich:1994cz}:
\begin{eqnarray} \label{rMp}
\bar M_P(\mu)^2= \bar M_P(0)^2-\frac{1}{96 \pi^2} \mu^2 N_l
\end{eqnarray}
with $N_l=N_S+N_\psi-4 N_V$ and where $\mu$ is the renormalization scale. The true energy scale  $\mu_*$ at which quantum gravity effects are large is one at which 
\begin{equation}
\label{strong}
\bar M^2 _P(\mu_*) \sim \mu_*^{2}.
\end{equation} 
This condition implies that fluctuations in spacetime geometry at length scales $\mu_*^{-1}$ will be unsuppressed. One finds
\begin{equation}
\label{strongsol}
\mu_*^{2}=  \frac{\bar M_P(0)^2}{1 + \frac{N_l}{96 \pi^2}}.
\end{equation} 

We can now trivially recalculate the amplitudes using our  renormalization group improvement and find that the energy scale $E^\star_{{\rm CM}}=\sqrt{s_\star}$ at which unitarity is violated  is given by 
\begin{eqnarray} 
E^\star_{{\rm CM}}=\bar M_P(E^\star_{{\rm CM}}) \sqrt{\frac{160 \pi}{N}}
\end{eqnarray}
where we have evaluated the Planck mass at the energy scale corresponding to the center of mass energy.  The new criteria for the consistency of the linearized theory is the following: If the scale at which gravity effects become strong is larger than the energy at which unitarity is violated, i.e. $\mu_\star > E^\star_{{\rm CM}}$ then linearized General Relativity  coupled  to the particle physics model under consideration is  inconsistent, on the other hand for  $\mu_\star \le E^\star_{{\rm CM}}$, the theory is well-behaved  up to energies $\mu_\star$ and the effective theory is consistent. This is our central result.  In terms of particle content, the criteria for the unitarity of the S-matrix up to the scale of strong gravity becomes
\begin{eqnarray}  \label{bound}
N=\frac{1}{3}N_S+N_\psi+4 N_V  \le 160 \pi. 
\end{eqnarray}
Note that the bound on $N$ is the same as the one obtained at tree level. However,  the requirement for a model to be consistent is  different in this new derivation. The two bounds coincide because we require that the true reduced Planck mass is the scale at which quantum gravitational effects become strong and not the Planck mass itself. Indeed, the reduced Planck mass appears in equation (\ref{strong}) and not the Planck mass. This is consistent with our previous observation that the expansion parameter for the metric is the reduced Planck mass and not the Planck mass.  The true reduced Planck mass itself depends on radiative corrections which turn out to be the same as those of the tree level scattering cross-sections. The requirement of having a solution to the equation (\ref{strong}) which fixes dynamically the true Planck mass leads to  the bound $N_l \ge -96 \pi^2$ which is weaker than the bound (\ref{bound}).  Models with a more negative $N_l$ do not lead to strong gravitational effects in which case gravity remains weak at all scales. 

\begin{figure}
\center
\includegraphics[scale=0.5]{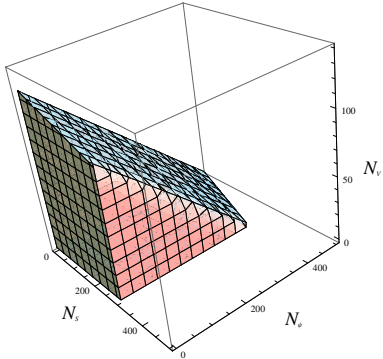}
\caption{$N_S$ is plotted on the $x$-axis while $N_\psi$ is plotted on the $y$-axis and $N_V$ is plotted on the $z$-axis. The parameter space that is compatible with unitarity of the S-matrix up to the true scale at which quantum gravity becomes strong is the colored region.}
\label{picbound}
\end{figure}

Using the same criteria as for the $J=2$ partial wave bound, one can obtain a second bound from the $J=0$ partial wave. This leads to a bound on the numbers of scalars:
\begin{eqnarray}  \label{bound2}
N_S  \le 96 \pi. 
\end{eqnarray}
We assumed that $\xi=0$, i.e., that the scalar fields are minimally coupled to gravity. As we shall see, this second bound turns out, in most cases, to be more restrictive than the $J=2$ bound for grand unified theories. The solution to the two inequalities (\ref{bound}) and (\ref{bound2})  is plotted in figure (\ref{picbound}). 

 In the standard model one has $N_S=4$, $N_\psi=45$ and $N_V=12$ and  one finds $N=283/3$,  $N_l=1$  which implies $E^\star_{{\rm CM}}= 2.3\ \bar M_P(0)$ and $\mu_\star \sim  \bar M_P(0)$. Linearized General Relativity coupled to the standard model is thus a valid effective theory up to the reduced Planck mass.  In the minimal supersymmetric Standard Model one has $N_S=98$, $N_\psi=61$ and $N_V=12$ and one finds $N=425/3$, $N_l=111$ and thus  $E^\star_{{\rm CM}}= 1.6\  \bar M_P(0)$ and $\mu_\star \sim 0.95 \ \bar M_P(0)$.  
One could be worried that in models with a larger particle content the theory could become inconsistent below the reduced Planck mass.  This strongly depends on the particle content. For example SO(10) with a $\bf 10$, $\bf 16$ and $\bf 45$ representations for the Higgs bosons leads to $N=781/3$, $N_S=97$ and $N_l=-35$. Our criteria implies that the corresponding effective theory is consistent. On the other hand in  grand unified SUSY SO(10) with Higgs bosons in the ${\bf 10}$, ${\bf 16}$, ${\bf \overline{16}}$ and ${\bf 770}$ representations one finds $N=4975/3$, $N_S=1720$, $N_l=2445$ and
 $E^\star_{{\rm CM}}= 0.41 \  \bar M_P(0)$ and $\mu_\star \sim  0.53 \ \bar M_P(0)$ which implies that the linearized effective theory is inconsistent. The same holds for the model proposed in \cite{Calmet:2008tn} where a hidden sector with $10^{32}$ particles of spin 0 and/or 1/2 leads to a reduced Planck mass at 1 TeV,  we find that unitarity is violated below this energy scale since both bounds are not fulfilled. Table 2 gives several more examples of models that pass or do not pass the tests. Note that supersymmetric models typically have more scalars and thus often face difficulties with unitarity below the scale at which quantum gravitational effects become strong. 
 
 \begin{table}[htb]
\begin{tabular}{|l||c|c|c||c|c|}
\hline
particle physics model & $N_l$ &$N$  &$N_S$ & {\rm J=2 bound} & {\rm J=0 bound} \\
\hline
\hline

standard model & $1$ & $283/3$ & $4$ & yes& yes\\
\hline
MSSM & $111$ & $425/3$ & $98$& yes& yes\\
\hline
\hline
SU(5) w/ ${\bf 5}$, ${\bf 24}$ & $-17$ & $457/3$ & $34$ & yes& yes\\
\hline
SU(5) w/ ${\bf 5}$, ${\bf 200}$ & $159$ & $211$ & $210$ & yes & yes \\
\hline
SU(5) w/ ${\bf 5}$, ${\bf 24}$, ${\bf 75}$ & $58$ & $532/3$ & $109$ & yes & yes  \\
\hline
SU(5) w/ ${\bf 5}$, ${\bf 24}$, ${\bf 75}$, ${\bf 200}$ & $258$ & $244$& $309$ &yes & no \\
\hline
SO(10) w/ ${\bf 10}$, ${\bf 16}$, ${\bf 45}$ & $-35$ & $781/3$ & 97 &yes & yes \\
\hline
SO(10) w/ ${\bf 10}$, ${\bf 16}$, ${\bf 210}$ & $130$ & $946/3$ & $262 $ & yes & yes\\
\hline
SO(10) w/ ${\bf 10}$, ${\bf 16}$, ${\bf 770}$ & $690$ & $502$ & $822$ & yes & no\\
\hline
\hline
SUSY-SU(5) w/ ${\bf 5}$, $\overline{\bf 5}$, ${\bf 24}$ & $165$ & $755/3$ & $158$ & yes& yes \\
\hline
SUSY-SU(5) w/ ${\bf 5}$, $\overline{\bf 5}$, ${\bf 24}$, ${\bf 75}$ & $390$ &  $1130/3$&$308$ & yes & no \\
\hline
SUSY-SU(5) w/ ${\bf 5}$, $\overline{\bf 5}$, ${\bf 200}$ & $693$ & $545$ & $510$ & no& no \\
\hline
SUSY-SO(10) w/ ${\bf 10}$, ${\bf 16}$, $\overline{\bf 16}$, ${\bf 45}$, ${\bf 54}$ & $432$ & $540$ & $378$ & no &no \\
\hline
SUSY-SO(10) w/ ${\bf 10}$, ${\bf 16}$, $\overline{\bf 16}$, ${\bf 210}$ & $765$ & $725$ & $600$ & no& no \\
\hline
SUSY-SO(10) w/ ${\bf 10}$, ${\bf 16}$, $\overline{\bf 16}$, ${\bf 770}$ & $2445$ & $4975/3$ & $1720$ & no & no \\
\hline
\hline
\end{tabular}
\caption{We consider different unification models which have been considered in the literature. Clearly models with large representations lead to unitarity problems. The last two columns are describing whether a given model passes the unitarity bound of the  $J=2$ ($N=\frac{1}{3}N_S+N_\psi+4 N_V  \le 160 \pi$) and $J=0$ ($N_S  \le 96 \pi$) partial waves.}
\end{table}
 
 Another extreme case of a model which suffers from a unitarity problem is the following. Consider a large hidden sector of particles  of spin 1 coupled to the standard model only gravitationally. The renormalization of the reduced Planck mass (\ref{rMp})  implies that the Planck mass increases with energy in that model while the scale at which unitarity is violated decreases with the number of spin 1 fields in the theory leading to a collapse of the effective theory description.  A caveat is that the renormalization of Newton's constant due to the graviton has not been included in our calculation. It is known that the graviton's contribution has the same sign as that of the spin 1 particle \cite{BjerrumBohr:2002ks}. It is however likely to be  a small effect. In particular, in models with a large number of fields, the graviton contribution to the renormalization of the Planck mass is a $1/N_l$ effect.  

Our results have interesting consequences for models of unification of General Relativity. For example in asymptotically safe gravity  \cite{fixedpoint}, the expansion parameter for the higher dimensional operators needs to be the reduced Planck mass $\sim \mu_\star$ and not the Planck mass:
 \begin{eqnarray} 
S[g]= -\int d^4x \sqrt{-\det(g)}\! \! \! \! \! && \left   (- \Lambda(\mu)^4+\frac{\bar M_P(\mu)^2}{32 \pi} R+  a(\mu) R_{\mu\nu} R^{\mu\nu} +b(\mu) R^2 
\right . \\ \nonumber && \left . +\frac{c(\mu)}{\mu_\star^2} R^3+\frac{d(\mu)}{\mu_\star^2} R R_{\mu\nu} R^{\mu\nu}+ .... \right )
\end{eqnarray}
where $\Lambda(\mu)$ is the running cosmological constant. This action reduces to  (\ref{efflag1}) once linearized. As we have seen the standard model would not lead to a consistent effective theory if one expanded into the Planck mass instead of the reduced Planck mass, as unitarity would be violated at about $1/2 M_P$.  Furthermore, this scenario is only viable for particle physics models with a particle content that fulfills the bound  (\ref{bound}),  since in that case linearized General Relativity is a valid effective theory up to the energy scale at which quantum gravity effects become relevant. 

In models that do not satisfy the bound  (\ref{bound}), there is, as we have seen, a violation of unitarity below the reduced Planck. A solution could come from embedding these models into string theory where the string scale appears as a new parameter. In these models, the Planck mass is related to the string scale $M_s$ and the extra-dimensional volume $V_6$ via the relation $M_P^2=1/g_s^2 M_s^8 V_6$ where $g_s$ is the string coupling constant. In this framework it is possible to decouple the gravitational scale from the string scale by adjusting the string coupling. An extreme example is that of little string theory in the TeV region \cite{antoniadis,calhsu} which requires $g_s \sim 10^{-16}$. If the string scale is lower than the Planck mass, non-local effects associated with the stringy nature of the particles could solve the unitarity problem. 

Finally note that the effect of the renormalization of the Planck mass also affects the bound we have obtained earlier for the coefficient of the non-minimal coupling of a scalar field to the Ricci scalar. Unitarity should hold up to $\mu_\star$ and not just $\bar M_P$. We thus find again that the unitarity limit is very sensitive to the particle content of the particle physics model.

{\it Conclusions:}  We have reconsidered the unitarity of the S-matrix for linearized General Relativity coupled to models of particle physics taking into account for the first time the renormalization of the Planck mass. We derive a bound on the particle content of the particle physics models coupled to General Relativity. Our results have significant implications for models trying to unify General Relativity with models of particle physics. In the case of the standard model and the minimal supersymmetric standard model and more generically in models satisfying our bound,  linearized General Relativity offers a theoretically consistent effective theory since there is no violation of unitarity below the reduced Planck mass which is the expansion parameter of the effective theory. If nature is described by one of these particle physics models, the fundamental theory of quantum gravity could be General Relativity itself which could be renormalizable at the non-perturbative level, i.e. asymptotically safe gravity, as proposed by Weinberg  some thirty years ago \cite{fixedpoint}.  In particle physics models  which do not satisfy the bound   (\ref{bound}), one  finds that the unitarity of the S-matrix is violated at an energy scale below the reduced Planck mass.  An extreme case example would be that of  asymptotically free gravity. Our results imply that asymptotically free gravity is  inconsistent. In less extreme cases, the cure could come from embedding models that do not satisfy the bound into a non local theory of quantum gravity.

\bigskip

{\it Acknowledgments:} 
We would like to thank Stephen Hsu and David Reeb for helpful discussions and valuable suggestions.
This work in supported in part by the European Cooperation in Science and Technology (COST) action MP0905 "Black Holes in a Violent Universe".


\bigskip


\end{document}